\documentstyle[prd,aps,preprint,epsfig]{revtex}
\begin{document}
%%%%%%%%%%%%%%%%%%%%%%%%%%%%%%%%
%\newfont\fiverm{cmr5}
%\input prepictex
%\input pictex
%\input postpictex
%\textheight 25 cm
\newcommand{\TeV}{\,{\rm TeV}}
\newcommand{\GeV}{\,{\rm GeV}}
\newcommand{\MeV}{\,{\rm MeV}}
\newcommand{\keV}{\,{\rm keV}}
\newcommand{\eV}{\,{\rm eV}}
\def\ap{\approx}
\def\bea{\begin{eqnarray}}
\def\eea{\end{eqnarray}}
\def\bec{\begin{center}}
\def\ec{\end{center}}
\def\pC{\tilde{\chi}^+}
\def\nC{\tilde{\chi}^-}
\def\pnC{\tilde{\chi}^{\pm}}
\def\Ne{\tilde{\chi}^0}
\def\snu{\tilde{\nu}}
\def\tN{\tilde N}
\def\ler{\lesssim}
\def\gtr{\gtrsim}
\def\beq{\begin{equation}}
\def\eeq{\end{equation}}
\def\haf{\frac{1}{2}}
\def\lpp{\lambda''}
\def\ccg{\cal G}
\def\slash#1{#1\!\!\!\!\!/}
\def\rpv{\slash{R_p}}
\def\ler{\lesssim}
\def\gtr{\gtrsim}
\def\pslash{p\hspace{-2.0mm}/}
\def\qslash{q\hspace{-2.0mm}/}
\def\p{\partial}
\def\f{\frac}

\newcommand{\imag}{\Im {\rm m}}
\newcommand{\real}{\Re {\rm e}}

%%%%% YW def %%%%%
\def\f#1#2{\frac{#1}{#2}}
\def\p{\partial}
\def\si{\sigma}
\def\Si{\Sigma}
\def\phivis{\Phi_{\rm vis}}
\def\phibulk{\Phi_{\rm bulk}}

\def\bea{\begin{eqnarray}}
\def\eea{\end{eqnarray}}
\def\beq{\begin{equation}}
\def\eeq{\end{equation}}
\setcounter{page}{1}
\preprint{KAIST-TH 03/01, KUNS-1821}
\title{SUSY Flavor Problem and
Warped Geometry}
\author{$^{(a)}$Kiwoon Choi,\footnote{kchoi@hep.kaist.ac.kr}
$^{(a)}$Do Young Kim,\footnote{kimdoyoung@hep.kaist.ac.kr}
$^{(a)}$Ian-Woo Kim,\footnote{iwkim@hep.kaist.ac.kr}
$^{(b)}$Tatsuo Kobayashi,\footnote{kobayash@gauge.scphys.kyoto-u.ac.jp}}
\address{$^{(a)}$Department of Physics,
Korea Advanced Institute of Science and Technology\\ Daejeon
305-701, Korea \\
$^{(b)}$ Department of Physics, Kyoto University,
Kyoto 606-8502, Japan}
\date{\today}
\maketitle
%%%%%%%%%%%%%%%%%%%%%%%%%%%%%%%%%%%%%%%%%%%%%%%%%%%%%%%%%%%%%%
\begin{abstract}
%%%%%%%%%%%%%%%%%%%%%%%%%%%%%%%%%%%%%%%%%%%%%%%%%%%%%%%%%%%%%%
We point out that
supersymmetric warped geometry can provide a solution to
the SUSY flavor problem, while generating
hierarchical Yukawa couplings.
In supersymmetric theories
in a slice of ${\rm AdS}_5$ with the Kaluza-Klein scale
$M_{KK}$ much higher than the weak scale, if all visible fields
originate from 5D bulk fields and  supersymmetry breaking is
mediated by the bulk radion superfield
and/or some brane chiral superfields,
potentially dangerous
soft scalar masses and trilinear $A$ parameters
at $M_{KK}$ can be naturally suppressed
compared to the gaugino masses by small warp factor.
We present simple models yielding
phenomenologically interesting patterns of soft parameters
in this framework.
%%%%%%%%%%%%%%%%%%%%%%%%%%%%%%%%%%%%%%%%%%%%%%%%%%%%%%%%%%%%%%
\end{abstract}
%%%%%%%%%%%%%%%%%%%%%%%%%%%%%%%%%%%%%%%%%%%%%%%%%%%%%%%%%%%%%%

%\section{Introduction}
\pagebreak

Warped extra dimension can provide a small geometric factor which
can be useful to explain various hierarchical structures in
particle physics. For instance, in 5-dimensional (5D) theory on a
slice of $\mbox{AdS}_5$ with AdS curvature $k$ and orbifold radius
$R$, the warp factor $e^{-\pi kR}$ may be responsible for the huge
hierarchy between the 4D Planck scale ($M_{Pl}\sim 10^{18}$ GeV)
and the weak scale ($M_W\sim 10^2$ GeV) \cite{Randall:1999ee},
and/or the hierarchical quark and lepton masses
\cite{Gherghetta:2000qt,Huber:2000ie}, and/or small neutrino
masses \cite{Grossman:1999ra}. In this paper, we wish to examine
the implications of warped geometry to the flavor structure of
soft SUSY breaking parameters as well as Yukawa couplings. 
To this end,
we consider supersymmetric 5D theory on a slice of $\mbox{AdS}_5$
in which all visible massless 4D modes arise from 5D bulk vector
multiplets and hypermultiplets \cite{Gherghetta:2000qt}. The
Kaluza-Klein (KK) scale $M_{KK}\approx k e^{-\pi kR}$ of the model
is assumed to be much higher than $M_W$, e.g. $10^{16}-10^{13}$
GeV, and then the stability of $M_W$ is achieved by the weak scale
supersymmetry (SUSY) as it is in the supersymmetric standard model. If
there is no light gauge-singlet 5D field
other than the minimal 5D supergravity (SUGRA) multiplet, 
SUSY breaking is mediated mainly by the radion superfield and/or 
some brane superfields
\footnote{The 4D SUGRA multiplet always participates in
the mediation of SUSY breaking through the conformal anomaly
\cite{anomaly}.
However as we will see, in the models under consideration
its contributions to soft parameters
are negligible compared to those from
the radion and brane superfields.}.
As we will see, in such case $\mbox{AdS}_5$ geometry
can naturally suppress the potentially dangerous
soft scalar masses and trilinear $A$
parameters at $M_{KK}$, while generating hierarchical Yukawa
couplings, for rather generic forms of flavor violation
in the underlying theory.
This would solve the SUSY flavor problem since the
squark and slepton masses at $M_W$
(at least of the first and second generations)
are generated mainly by the
flavor-independent renormalization group (RG) evolution arising 
from the unsuppressed gaugino masses.
Also this framework using the warped geometry 
to solve the SUSY flavor problem
can provide predictions for the shape of soft
parameters at $M_{KK}$, which can lead to interesting
phenomenology at $M_W$.

Our starting point is generic 5D gauge theory
coupled to the minimal 5D SUGRA on
$S^1/Z_2$. The action of the model is given by
\cite{Altendorfer:2000rr,Gherghetta:2000qt} \bea \label{5daction1}
S = \int d^5x&&\sqrt{-G} \,\left[\frac{1}{2}\left(\, {\cal
R}+\bar{\Psi}^i_M\gamma^{MNP}D_N\Psi_{iP}-\frac{3}{2}C_{MN}C^{MN}
%-\frac{3}{2}k\epsilon(y)\bar{\Psi}^i_M\gamma^{MN}\Psi_{iN}+
+12k^2\,\right) \right.\nonumber \\
&&+\frac{1}{{g}_{5a}^2}
\left(-\frac{1}{4}F^{aMN}F^a_{MN}
+\frac{1}{2}D_M\phi^aD^M\phi^a
+\frac{i}{2}\bar{\lambda}^{ia}\gamma^MD_M\lambda^a_i
%+\frac{1}{2}k\epsilon(y)\bar{\lambda}^{ia}\lambda^a_i-
%4k^2\phi^a\phi^a
\right)\nonumber \\
&&\left.+|D_Mh_I^i|^2+i\bar{\Psi}_I\gamma^MD_M\Psi_I+
i\tilde{c}_Ik\epsilon(y)\bar{\Psi}_I\Psi_I +... \right]
\eea
where ${\cal R}$ is the 5D Ricci scalar, $\Psi^i_M$ ($i=1,2$) are
the symplectic Majorana gravitinos,
$C_{MN}=\partial_MB_N-\partial_NB_M$ is the graviphoton field strength, and
$y$ is the 5th coordinate with fundamental range
$0\leq y\leq\pi$.
Here $\phi^a,A_M^a$ and $\lambda^{ia}$ are 5D scalar, vector and
symplectic Majorana
spinors constituting a 5D vector multiplet,
$h_I^i$ and $\Psi_I$  are 5D scalar and Dirac spinor constituting
the $I$-th hypermultiplet
with kink mass $\tilde{c}_Ik\epsilon(y)$.
Note that we set the 5D Planck mass
$M_5=1$ and all dimensionful parameters, e.g.
the 5D gauge coupling $g_{5a}$ and the AdS curvature
$k$, are defined in this unit.
With appropriate values of the brane cosmological constants
at the orbifold fixed points ($y=0,\pi$), the ground state geometry
is given by a slice of ${\rm AdS}_5$
having the radion ($R$) dependence:
\beq
\label{metric}
G_{MN}dx^Mdx^N=e^{-2kRy}\eta_{\mu\nu}dx^\mu dx^\nu+R^2dy^2.
\eeq

It is convenient to write the above 5D action in $N=1$
superspace\cite{Arkani-Hamed:2001tb,Marti:2001iw,Linch:2002wg}.
For the 5D SUGRA multiplet, we keep only the radion superfield
$$
T=\left(\,R+iB_5\,, \,\,\frac{1}{2}(1+\gamma_5)\Psi^{i=2}_5\,\right)\,
$$
and replace other fields by their vacuum expectation values. For
the 5D vector multiplets and hypermultiplets, one needs
appropriate $R$ and $B_5$-dependent field redefinitions to
construct the corresponding $N=1$ superfields
\cite{Marti:2001iw,Choi:2002wx}. After such field redefinition,
the relevant piece of the action is given by\footnote{
Note that our superfield basis for hypermultiplets differs from
\cite{Marti:2001iw} which is related to ours
by $H_I\rightarrow e^{-(\frac{3}{2}-\tilde{c}_I)Tk|y|}H_I$
and $H^c_I\rightarrow e^{-(\frac{3}{2}+\tilde{c}_I)Tk|y|}H^c_I$.}
\cite{Arkani-Hamed:2001tb,Marti:2001iw} \bea \label{5daction2}
&&\int d^5x \,\left[\,\int d^4\theta
\,\left\{\frac{T+T^*}{2}\left(
e^{(\frac{1}{2}-\tilde{c}_I)(T+T^*)k|y|}H_IH_I^*
+e^{(\frac{1}{2}+\tilde{c}_I)(T+T^*)k|y|}H^c_IH^{c*}_I\right)
\right\}\right. \nonumber \\ && \quad\quad\quad\left. +\left\{\,
\int d^2\theta\, \frac{1}{4{g}_{5a}^2}TW^{a\alpha}W^a_{\alpha}
+h.c.\,\right\} \right]\,,\eea where $W^{a}_{\alpha}$ is the
chiral spinor superfield for the vector superfield ${\cal V}^a$
containing $(A^a_{\mu},\lambda^a)$ with
$\lambda^a=\frac{1}{2}(1-\gamma_5)\lambda^{a1}$, $H_I$ and $H_I^c$
are chiral superfields containing $(h_I^1,\psi_I)$ and
$(h_I^{2*},\psi^c_I)$, respectively, with
$\psi_I=\frac{1}{2}(1-\gamma_5)\Psi_I$,
$\bar{\psi^c}_I=\frac{1}{2}(1+\gamma_5)\Psi_I$. As the theory is
orbifolded by $Z_2: y\rightarrow -y$, all 5D fields should have a
definite boundary condition under $Z_2$.
%\footnote{The generalization of our discussion to
%$S^1/Z_2\times Z_2^\prime$ where $Z_2^\prime:
%\pi+y\rightarrow \pi-y$ is straightforward \cite{choi}.}
The 5D SUGRA multiplet is assumed to have the standard
boundary condition
leaving the 4D $N=1$ SUSY unbroken.
As for ${\cal V}^a$, one needs
${\cal V}^a(-y)={\cal V}^a(y)$ to obtain massless 4D gauge multiplet.
On the other hand,
$H_I (-y) = z_I H_I (y)$ and
$H_I^c(-y) = -z_I H_I (y)$ ($z_I=\pm 1$), and then massless
4D matter multiplet can originate either from
$H_I$ or $H^c_I$ depending on the sign of $z_I$.

In addition to the bulk action (\ref{5daction1}),
there can be brane actions at the fixed points
$y=0,\pi$.
The general covariance requires that
the 4D metric in brane action should be the
4D component of the 5D metric at fixed point.
Using the general covariance and also
the $R$ and $B_5$-dependent field redefinitions
which have been made to construct $N=1$ superfields,
one can easily find
the $T$-dependence of brane actions \cite{Marti:2001iw,Casas,Hebecker}.
For instance,
the brane actions which would be relevant for
Yukawa couplings and soft parameters
are given by\footnote{
The chiral anomaly of the $R$ and $B_5$-dependent field redefinition
induces $T$-dependent pieces in
$\omega_a$ and $\omega^\prime_a$ \cite{Choi:2002wx}.
But they are loop-suppressed and
not very relevant for the discussion in this paper.}
\bea
\label{braneaction}
S_{\rm brane}&=&\int\,d^5x\,
\left[\,\,\delta(y)\,\left\{
\int d^4\theta \,L_{IJ}(Z,Z^*)\tilde{H}_I\tilde{H}^*_J
\right.\right.\nonumber \\
&& \left.+\left(\,\int d^2\theta
\left(
\,\frac{1}{4} \omega_a(Z) W^{a\alpha}W^a_{\alpha}
+\lambda_{IJK}(Z)\tilde{H}_I\tilde{H}_J\tilde{H}_K\right)
+{\rm h.c}\,\right)\,\right\}
\nonumber \\
&&+\,\,\delta(y-\pi)\left\{
\int d^4\theta \, e^{-(c_I\pi kT+c_J\pi kT^*)}
L^\prime_{IJ}(Z^\prime,Z^{\prime *})\tilde{H}_I\tilde{H}^{*}_J
\right.
\nonumber \\
&&\left.\left.+\left(\,\int d^2\theta 
\left(\,\frac{1}{4}
\omega_a^\prime(Z^\prime)
W^{a\alpha}W^a_{\alpha}
+e^{-(c_I+c_J+c_K)\pi kT}\lambda^\prime_{IJK}(Z^\prime)
\tilde{H}_I\tilde{H}_J\tilde{H}_K
\right)
+{\rm h.c}\,\right)\,\right\}\,\right]
\eea
where $Z$ and $Z^\prime$ denote generic 4D
chiral superfields living {\it only} on the brane at $y=0$ and $y=\pi$,
respectively,
$\tilde{H}_I=H_I$ for $z_I=1$, while
$\tilde{H}_I=H_I^c$ and
for $z_I=-1$, and
$$
c_I=z_I\tilde{c}_I-\frac{1}{2}.
$$
Here $L_{IJ}$ ($L_{IJ}^\prime$) are generic hermitian functions
of $Z$ and $Z^*$ ($Z^\prime$ and $Z^{\prime *}$),
and $\omega_a$ and $\lambda_{IJK}$
($\omega^\prime$ and $\lambda^\prime_{IJK}$) are
generic holomorphic functions of $Z$ ($Z^\prime$).

The 4D Yukawa couplings and soft parameters can be
studied by constructing the effective action
of massless 4D superfields.
In our superfield basis, the {\it $y$-independent} modes of
5D superfields correspond to massless 4D superfields.
Let $V^a$ denote the $y$-independent mode of
${\cal V}^a$, and
$Q_I$ to be the $y$-independent mode
of $H_I$ when $z_I=1$ or of $H_I^c$ when $z_I=
-1$. Here we will assume that all
visible 4D gauge and matter fields are in $\{V^a,Q_I\}$,
and examine their Yukawa couplings and soft SUSY breaking
parameters when the SUSY breaking is mediated
by the brane superfields $Z,Z^\prime$ and the radion superfield $T$.
Those Yukawa couplings and soft parameters
at the Kaluza-Klein(KK) scale
$M_{KK}\approx ke^{-\pi kR}$ can be evaluated from
the 4D effective action which can be written as
\bea
\label{4deffective1}
\left[\,\int d^4 \theta\,\, Y_{IJ}Q_IQ^*_J\,\right]
+\left[\,\int d^2 \theta
\,\left(\f{1}{4} f_a W^{a\alpha} W^a_\alpha + \tilde{y}_{IJK}Q_IQ_JQ_K\,\right)
 +\mbox{h.c.}\,
\right],
\eea
where $Y_{IJ}$ are hermitian wave function coefficients,
$f_a$ are holomorphic gauge kinetic functions,
and $\tilde{y}_{IJK}$ are holomorphic Yukawa couplings.
From (\ref{5daction2}) and (\ref{braneaction}),
we find
\bea
\label{4deffective}
Y_{IJ}&=&\frac{1}{{c}_Ik}\left( 1- e^{-{c}_I\pi k(T+T^*)}
\right)\delta_{IJ}+L_{IJ}(Z,Z^*)
+\frac{L^\prime_{IJ}(Z^\prime,Z^{\prime *})
}{e^{({c}_I\pi kT+{c}_J\pi kT^*)}},
\nonumber \\
f_a&=&\frac{2\pi}{g_{5a}^2}T+\omega_a(Z)+\omega^\prime_a(Z^\prime),
\quad
\tilde{y}_{IJK}=\lambda_{IJK}(Z)+\frac{\lambda^\prime_{IJK}(Z^\prime)}{
e^{({c}_I+{c}_J+{c}_K)\pi kT}}. \eea Note that 5D SUSY in bulk
enforces that the Yukawa couplings of $Q_I$ originate entirely
from the brane action (\ref{braneaction}).

It is straightforward to compute soft parameters for
generic forms of $L_{IJ}, L^\prime_{IJ}$.
Here we will focus on
\bea
L_{IJ}(Z,Z^*)=-\kappa_{IJ}ZZ^*,\quad
L^\prime_{IJ}(Z^\prime,Z^{\prime *})=-\kappa^\prime_{IJ}
 Z^\prime Z^{\prime *},
\eea where $\kappa_{IJ}$ and $\kappa^\prime_{IJ}$ are generic
constants of order one, while keeping $\omega_a$ and $\lambda_{IJK}$
($\omega^\prime_a$ and $\lambda^\prime_{IJK}$) as generic
holomorphic functions of $Z$ ($Z^\prime$). The results for generic
forms of $L_{IJ}$ and $L^\prime_{IJ}$ will be presented elsewhere
\cite{futurework}. In regard to
the suppression of soft scalar masses and trilinear $A$ parameters
at $M_{KK}$, those generic results show the same behavior as
our case. 
For simplicity, we further assume that $\langle Z\rangle\ll 1$ and
$\langle Z^\prime \rangle \ll 1$ (in the unit with $M_5=1$), so
$\langle L_{IJ}\rangle \ll 1$ and $\langle L^\prime_{IJ}
\rangle\ll 1$ though $\kappa_{IJ}$ and $\kappa^\prime_{IJ}$ are
generically of order one. However, in regard to SUSY breaking,
we consider the most general case in which 
any of the $F$-components of
$T,Z$ and $Z^\prime$ can be the major source of SUSY breaking. 
Note that here we are not concerned with
the dynamial origin of those $F$-components,
but with the resulting soft
parameters of visible fields for generic values of
the $F$-components.

The Yukawa couplings  $y_{IJK}\phi^I\psi^J\psi^K$
of {\it canonically normalized}
superfield $Q_I=\phi^I+\theta \psi^I+\theta^2 F^I$
are easily obtained from (\ref{4deffective}):
\bea
\label{yukawa}
y_{IJK}=(Y_IY_JY_K)^{-1/2}
\left(\lambda_{IJK}+\frac{\lambda^\prime_{IJK}}{e^{(c_I+c_J+c_K)\pi kT}}
\right),
\eea
where $Y_I=(1-e^{-c_I\pi k(T+T^*)})/kc_I$.
The gaugino masses
$\frac{1}{2}M_a\lambda^a\lambda^a$ for canonically
normalized gauginos $\lambda^a$,
the soft scalar mass-squares $m^2_{I\bar{J}}\phi^I\phi^{J*}$,
and the trilinear $A$-terms $A_{IJK}\phi^I\phi^J\phi^K$ for canonically
normalized scalar fields $\phi^I$ are also obtained to be
\bea
\label{soft}
M_a &=& \frac{1}{2}g_a^2 \left(
\frac{2\pi}{g_{5a}^2}F^T+\frac{\partial \omega_a}{\partial Z}
F^Z+\frac{\partial \omega^\prime_a}{\partial Z^\prime}
F^{Z^\prime}\right)\,,
\nonumber \\
m^2_{I\bar{J}} &=& (Y_IY_J)^{-1/2}\left[
\,\frac{\pi^2 c_Ik\delta_{IJ}|F^T|^2}{e^{c_I\pi k(T+T^*)}-1}
+\kappa_{IJ}\left| F^Z \right|^2
+\frac{\kappa^\prime_{IJ}|F^{Z^\prime}|^2}{e^{(c_I\pi
kT+c_J\pi kT^*)}}\,\right],
\nonumber \\
A_{IJK} &=& (Y_IY_JY_K)^{-1/2}\left[F^T
\frac{\partial}{\partial T}\ln
\left(\frac{\lambda_{IJK}+\lambda^\prime_{IJK}e^{-(c_I+c_J+c_K)\pi kT}}{
Y_IY_JY_K}\right)\right.
\nonumber \\
&&\left. \times \left(\lambda_{IJK}+\frac{\lambda^\prime_{IJK}}{
e^{(c_I+c_J+c_K)\pi kT}}\right)
+F^Z\frac{\partial \lambda_{IJK}}{\partial Z}
+\frac{F^{Z^\prime}}{e^{(c_I+c_J+c_K)\pi kT}}\frac{\partial
\lambda^\prime_{IJK}}{\partial Z^\prime}\,\right]\,,
\eea
where $g_a^2$ are 4D gauge couplings,
and  $F^T,F^Z$ and $F^{Z^\prime}$ denote the $F$-components
of $T,Z$ and $Z^\prime$, respectively.

One can now consider two simple models
in which the SUSY flavor
problem is solved (or ameliorated) by warped geometry.
Here we will briefly describe the models, while leaving
the details of the models including the phenomenological
aspects to the subsequent work \cite{futurework}.
In the model (I), 
Yukawa couplings exist {\it only} at $y=0$,
i.e. $\lambda^\prime_{IJK}=0$, and 
there is {\it no} SUSY breaking $Z$ at $y=0$.
Then an appropriate radion stabilization mechanism
is assumed to yield $e^{-\pi kR}\approx 10^{-2}-10^{-5}$
for which $M_{KK}\approx ke^{-\pi kR}\approx 10^{16}-10^{13}$ GeV.
Note that these conditions are stable against radiative corrections.
We further assume that $c_I=0$ for Higgs superfields
$Q_I$, while $c_{J,K}> 0$ with $e^{-c_J\pi kR}\ll 1$ and
$e^{-c_K\pi kR}\ll 1$ for $Q_J$ and $Q_K$ denoting
the quark and/or lepton superfields.
In this model, the quark and lepton Yukawa couplings
are given by
\beq
\label{yukawa1}
y_{IJK}=\left(\frac{k^2 c_Jc_K}{2\pi R}\right)^{1/2} \lambda_{IJK},
\eeq
so warped geometry is not responsible for
hierarchical Yukawa couplings.
As for the soft parameters at $M_{KK}$, $M_a={\cal O}(F^T)$
and/or ${\cal O}(F^{Z^\prime})$, and 
\bea
\label{soft1}
&&m^2_{J\bar{K}}
=\frac{k(c_Jc_K)^{1/2}}{e^{(c_J+c_K)\pi kR}}
\left(\,\pi^2kc_J\delta_{JK}\left|
F^T\right|^2+\kappa^\prime_{JK}\left|
F^{Z^\prime}\right|^2\,\right)
={\cal O}(e^{-(c_J+c_K)\pi kR}M_a^2),
\nonumber \\
&&A_{IJK}=-y_{IJK}\frac{F^T}{2R}={\cal O}(y_{IJK}M_a).
\eea
Note that the squark and slepton
mass-squares $m^2_{J\bar{K}}$  are suppressed by
small warp factor. On the other hand,
$A_{IJK}/y_{IJK}$ is unsuppressed, but it is {\it universal}
and generically contains nonzero CP phase.
As a result, $m^2_{J\bar{K}}$ of
the first and second generations {\it at the weak scale}
are (approximately) flavor-independent
as they are generated mainly
by the RG evolution arising from $M_a$.
In fact, one can consider the variation of the model (I)
in which $e^{-c_I\pi kR}\ll 1$ for
Higgs superfields $Q_I$. In such  model,
both $A_{IJK}/y_{IJK}$ and $m_{J\bar{K}}$ at $M_{KK}$
are exponentially suppressed compared
to $M_a$.

In the model (II),
$\lambda_{IJK}=0$ and
all $\lambda^\prime_{IJK}$ are of order unity,
and there is {\it no} SUSY breaking $Z$ at $y=0$.
The radion is stabilized again at $e^{-\pi kR}=10^{-2}-10^{-5}$,
and we assume that $c_I=0$ for Higgs superfield
$Q_I$, while $c_{J,K}\geq 0$ for quark and/or lepton
superfields $Q_{J,K}$.
In this model, hierachical Yukawa couplings are naturally
generated by warped geometry:
\bea
\label{yukawa2}
y_{IJK}=\left(\frac{1}{(2\pi R Y_JY_K)^{1/2}}\right)
\frac{\lambda^\prime_{IJK}}{e^{(c_J+c_K)\pi kR}}
={\cal O}(e^{-(c_J+c_K)\pi kR}),
\eea
where $Y_J=(1-e^{-2c_J\pi kR})/kc_J=2\pi R$ when $c_J=0$,
and $Y_J=1/kc_J$ when $e^{-c_J\pi kR}\ll 1$.
Warped geometry similarly suppresses
$m^2_{J\bar{K}}$ and
$A_{IJK}$ at $M_{KK}$ as
\bea
\label{soft2}
 m^2_{J\bar{K}}&=&
(Y_JY_K)^{-1/2}\left(
\frac{\pi^2 c_Jk\delta_{JK}|F^T|^2}{e^{2c_J\pi kR}-1}
+\frac{\kappa^\prime_{JK}|F^{Z^\prime}|^2}{e^{(c_J+c_K)\pi kR}}
\right)
={\cal O}(e^{-(c_J+c_K)\pi kR}M_a^2),
\nonumber \\
A_{IJK}&=&\frac{e^{-(c_J+c_K)\pi kR}}{(2\pi RY_JY_K)^{1/2}}
\left[-\frac{F^T}{2R}
\left(1+2(c_J+c_K)\pi kR+
\frac{2c_J\pi kR}{e^{2c_J\pi kR}-1}\right.\right.
\nonumber \\
&&\left.\left.+\frac{2c_K\pi kR}{e^{2c_K\pi kR}-1}\right)\lambda^\prime_{IJK}
+F^{Z^\prime}\frac{\partial\lambda^\prime_{IJK}}{\partial
Z^\prime}\right]=
{\cal O}(y_{IJK} M_a),
\eea
where again $M_a={\cal O}(F^T)$ and/or ${\cal O}(F^{Z^\prime})$.
An interesting feature of this model is that
$m^2_{J\bar{K}}$ at $M_{KK}$
have a correlation with Yukawa couplings.
To be specific, let us consider the case
with $n_{q_i}=(3,2,0)$,
$n_{u_i}=
(5,2,0)$, $n_{d_i}=(2,1,1)$ where  $n_K=-c_K\pi kR/\ln(0.2)$
for the 3-generations
of $SU(2)$-doublet quark superfields
$q_i$, and the $SU(2)$-singlet antiquark superfields $u_i,d_i$.
This model  gives
the up-quark Yukawa couplings $y^u_{ij}\sim
0.2^{n_{q_i}+n_{u_j}}$ and
the down-quark Yukawa couplings $y^d_{ij}
\sim 0.2^{n_{q_i}+n_{d_j}}$,
producing the correct quark masses and mixing angles.
At the same time, the model predicts the following patterns
of squark mass-squares
at $M_{KK}$:  $m^2_{ij}(\tilde{q})/M_a^2\sim
0.2^{n_{q_i}+n_{q_j}}$,
$m^2_{ij}(\tilde{u})/M_a^2\sim 0.2^{n_{u_i}+n_{u_j}}$,
and $m^2_{ij}(\tilde{d})/M_a^2\sim 0.2^{n_{d_i}
+n_{d_j}}$. So in the model (II) also,
the squark masses at the weak scale
of the first and second generations are
generated mainly by the flavor-independent RG evolution
arising from $M_a$.
Note that this model can give a sizable $23$-component of
the squark mass-square matrices, which may
lead to interesting phenomenology \cite{futurework}.

When $c_{J,K}\leq 0$ for the quark and lepton superfields
$Q_{J,K}$, one can obtain similar models yielding 
exponentially suppressed Yukawa couplings and/or soft
parameters by  exchanging $(Z, \kappa_{JK},
\lambda_{IJK})$ and
$(Z^\prime,\kappa^\prime_{JK},\lambda^\prime_{IJK})$,
in the previously described models for $c_{J,K}\geq 0$. 
At any rate, the suppression of
Yukawa couplings and soft parameters by warp factor has a  simple
geometric interpretation.
If $e^{-c_J\pi kR}\ll 1$ ($e^{-c_J\pi
kR}\gg 1$), the wavefunction of $Q_J$ is localized near at $y=0$
($y=\pi)$ with an exponentially small tale at $y=\pi$
($y=0$).
As a consequence, its overlaps with the brane Yukawa coupling
$\lambda^\prime_{IJK}$ ($\lambda_{IJK}$) and the brane SUSY
breaking $F^{Z^\prime}$ ($F^Z$) at $y=\pi$ ($y=0$) are
exponentially suppressed.
Also the wavefunction coefficients $Y_{JK}$ of such $Q_{J,K}$
are exponentially insensitive to the distance between two fixed points,
so to the value of $T$,
explaining why the contributions of $F^T$
to $m^2_{J\bar{K}}$ are exponentially suppressed independently
of the sign of $c_{J,K}$\footnote{
The suppression of the $F^T$-contribution to $m^2_{J\bar{K}}$
has been interpreted in \cite{Marti:2001iw} by means of
the AdS/CFT correspondence.}.
On the other hand, the wavefunction of
$V^a$ is constant over the 5-th dimension, so there is no such
suppression in gaugino masses. 

The results (\ref{yukawa1})-(\ref{soft2}) are obtained 
for the 4D effective action (\ref{4deffective1}) without including
the possible threshold corrections due to massive
KK modes.
Obviously, the (exponential) suppressions
of $y_{IJK}$ and $A_{IJK}$ 
are  stable against KK threshold corrections as they are due to 
the (exponentially) small holomorphic Yukawa couplings $\tilde{y}_{IJK}$.
On the other hand, $m^2_{J\bar{K}}$ generically get
threshold corrections of order $M_a^2/8\pi^2$.
In fact,
the structure of flavor-violating interactions
in the models (I) and (II) 
suggests that the flavor-violating part of the KK threshold corrections
to $m^2_{J\bar{K}}$ is further suppressed by
a small factor involving $\lambda_{IJK}$ or
$\lambda^\prime_{IJK}/e^{(c_J+c_K)\pi kR}$
or $\kappa^\prime_{JK}/e^{(c_J+c_K)\pi kR}$ \cite{futurework}.
There can be additional corrections to soft parameters
which are induced by non-renormalizable SUGRA interactions 
in 4D effective action \cite{choi}, 
but they are suppressed by
$M_{KK}^2/8\pi^2 M_{Pl}^2\sim e^{-2\pi kR}/8\pi^2$, so
are small enough.
There are also the model-independent SUSY breaking effects mediated
by the 4D superconformal  anomaly \cite{anomaly}.
In the models (I) and (II), we obtain the gravitino mass $m_{3/2}=
{\cal O}(e^{-\pi kR} M_a)$, so
the anomaly-mediated contributions to soft parameters are
$\delta M_a\sim e^{-\pi kR}M_a/8\pi^2$,
$\delta m^2_{J\bar{K}}\sim e^{-2\pi kR}M^2_a/(8\pi^2)^2$ 
and $\delta A_{IJK}\sim e^{-\pi kR}M_a/8\pi^2$, which
are small enough to be ignored.
So the leading radiative corrections to 
Yukawa couplings and soft parameters at $M_W$ in the
models (I) and (II) come from
the standard RG running down to $M_W$
starting from the boundary values at $M_{KK}$ given by
(\ref{yukawa1})-(\ref{soft2}).
%As we have noted, such radiative corrections are 
%responsible for the masses of
%the first and second generations of squarks and sleptons at $M_W$.

We finally note that the idea of AdS/CFT correspondence suggests 
a CFT framework which would reproduce the main
features  of our AdS models.
Indeed, models involving superconformal (SC) sector have been proposed
to generate hierarchical Yukawa couplings as well as
exponentially suppressed soft masses
\cite{Nelson:2000sn,Luty:2001jh}.
It is unclear yet the possible connection between
these SC models and the models (I) and (II) presented here, 
though one can easily find
the correspondence: $c_I\rightarrow \gamma_I/2$ and
$\pi kR\rightarrow \ln (\Lambda/M_c)$, where
$\gamma_I$ is the anomalous dimension of $Q_I$
driven by the coupling to the SC sector,
and $\Lambda$ and $M_c$ are the cutoff scale and the decoupling scale
of the SC sector, respectively.
At any rate, the AdS approach discussed in this paper
provides interesting perturbative framework to solve 
the SUSY flavor problem and the Yukawa hierarchy problem.

To conclude, we have noted that supersymmetric 5D theory on a slice
of $\mbox{AdS}_5$ with the Kaluza-Klein scale
$M_{KK}\approx 10^{16}-10^{13}$ GeV
can provide a solution to the
SUSY flavor problem, while generating hierarchical Yukawa
couplings. This framework utilizes the AdS warp factor
$e^{-\pi kR}\approx 10^{-2}-10^{-5}$ to
suppress the soft scalar masses and trilinear $A$-parameters
at $M_{KK}$, and provides 
phenomenologically interesting prediction for
the patterns of soft parameters.

\bigskip
\noindent
{\bf Acknowledgments}

KC, DYK and IWK are supported by KRF PBRG 2002-070-C00022, 
KRF Grant 2000-015-DP0080, the KOSEF Sundo Grant, and the 
Center for High Energy Physics of Kyungbook National University,
and TK is supported by the Grants-in-Aid for the Promotion
of Science No. 14540252.

\end{document}